\documentclass[aps,preprintnumbers,showpacs,twocolumn,superscriptaddress,floatfix,nofootinbib]{revtex4}

\usepackage[dvips]{graphicx}
\usepackage{times}
\usepackage{latexsym}
\usepackage{amssymb}
\usepackage{wasysym}
\usepackage{color}

\newcommand{\ie}{\textit{i.e.}~}
\newcommand{\eg}{\textit{e.g.}~}
\newcommand{\cf}{\textit{cf.}~}

\begin{document}

\title{On the ergoregion instability in rotating gravastars}

\author{Cecilia B. M. H. Chirenti}
\affiliation{
Max-Planck-Institut f\"ur Gravitationsphysik, 
Albert Einstein Institut,  
14476 Golm, Germany
}

\author{Luciano Rezzolla}
\affiliation{
Max-Planck-Institut f\"ur Gravitationsphysik, 
Albert Einstein Institut, 
14476 Golm, Germany
}

\affiliation{
Department of Physics, Louisiana State University, 
Baton Rouge, LA 70803 USA
}

\affiliation{
INFN, Department of Physics, University of Trieste, Trieste, Italy
}

\date{\today}

\begin{abstract}
The ergoregion instability is known to affect very compact objects
that rotate very rapidly and do not possess a horizon. We present here
a detailed analysis on the relevance of the ergoregion instability for
the viability of gravastars. Expanding on some recent results, we show
that not all rotating gravastars are unstable. Rather, stable models
can be constructed also with $J/M^2 \sim 1$, where $J$ and $M$ are the
angular momentum and mass of the gravastar, respectively. The genesis
of gravastars is still highly speculative and fundamentally unclear if
not dubious. Yet, their existence cannot be ruled out by invoking the
ergoregion instability. For the same reason, not all ultra-compact
astrophysical objects rotating with $J/M^2 \lesssim 1$ are to be
considered necessarily black holes.
\end{abstract}

\pacs{04.40.Dg, 04.30.Nk, 04.25.Nx}
\maketitle

\section{Introduction}

Gravastars have been recently presented by Mazur and
Mottola~\cite{Mazur} as a new exact solution to the Einstein equations. In
the original suggestion these are compact, spherically symmetric and
non-singular objects, that can be taken to be almost as compact as
black holes. In this ``three-layer'' model, the high compactness is
supported by a de-Sitter core, surrounded by a shell of matter, and the
exterior vacuum spacetime is, of course, that of the Schwarzschild
solution.

According to the original picture proposed in ref.~\cite{Mazur}, a
massive collapsing star could go through a phase transition when its
radius approaches $R = 2M$, forming a gravastar instead of a black
hole.  Although the dynamical processes that would lead to the
formation of a gravastar are far from being understood and could
probably be described as rather exotic, the final state can be
described by an exact (and fairly simple) solution of the Einstein
equations. This is what stimulates our interest in this matter.

The original gravastar model has inspired many subsequent works, but
always describing stationary solutions 
(except from ref.~\cite{Cardoso}, as we will see below). 
A related simplified model with an infinitesimally thin shell was
proposed in ref.~\cite{Visser} and later also generalized in
ref.~\cite{Carter}. Several 
possibilities for the interior solution have been considered in 
refs.~\cite{Bilic}, \cite{Lobo1} and \cite{Lobo2}, among others. More
recently, a solution for electrically charged gravastar configurations
was proposed in ref.~\cite{Horvat} and limits on the existence of
gravastars from astronomical data were considered in ref.~\cite{Broderick}.

Also an effort has been made to assess the properties of gravastars
when these are perturbed~\cite{DeBenedictis,Chirenti}. The result of
this analysis has lead, among other things, to the evidence that
perturbations in spherical gravastars can be used to discriminate them
from black holes, thus removing one of the most serious consequences
of the existence of such \textit{ultra-compact}
objects~\cite{Chirenti}. Note that hereafter we define as
{ultra-compact} any stellar object whose compactness $\mu\equiv M/R$
is {much} larger than that of typical neutron stars of comparable mass
and angular momentum, and which is $\mu \sim 0.15-0.2$. This
definition is inevitably weak and ambiguous, but it aims at focussing
on the large compactness of gravastars as the most relevant
property. Clearly, black holes are always more compact than any
possible gravastar model (although their compactness is
only infinitesimally smaller for nonrotating models) but the relevant
point to bear in mind is that gravastars have compactness much larger
than that of standard stars and comparable (although smaller) to that
of black holes.

Further expanding on the perturbative analysis carried out in
ref.~\cite{Chirenti}, Cardoso et al.~\cite{Cardoso} have recently
considered the properties of perturbed and rotating gravastars and
assessed, within the slow-rotation approximation, their stability
against the \textit{``ergoregion instability''}. We recall that such
instability affects rapidly rotating and very compact stellar objects
which do have an ergoregion but do not have an event
horizon~\cite{Comins,Yoshida}.

The Kerr black hole is a good example of an object that has an
ergoregion. But an ergoregion can also develop in compact stars that are
sufficiently rapidly rotating. In this region,  the relativistic frame
dragging is so strong that no 
stationary orbits are allowed. All trajectories of particles in this
region must rotate in the same direction of the rotation of the star. 

Because of this effect,
some particles in the ergoregion can be measured by an observer at
rest at
infinity as having negative energy. This happens because this observer
measures the energy of the particles by projecting their four-momentum
vectors onto his four-velocity. As the observer is stationary (and no
stationary trajectories are allowed inside the ergoregion), his
four-velocity is outside of the light cone of the particles in the
ergoregion. This causes some of these particles 
to have their energy measured as negative by the observer at infinity. 
 
The instability occurs then in the following way.
From an initially small perturbation with negative energy trapped in the
ergoregion, one can extract positive energy (that leaves the star and
goes to infinity) by increasing the negative energy inside the
ergoregion (thus conserving the total energy). As the negative energy
trapped in the ergoregion increases, the star radiates even more
positive energy to infinity and this process leads to the
instability. This process is very general, and scalar, electromagnetic
and gravitational waves become unstable in a star with an ergoregion.

This paper is dedicated to reconsider the analysis carried out by
Cardoso et al.~\cite{Cardoso} and to extend it to a larger space of
possible models, taking into account the limits on the thickness of
the matter shell and on its compactness. When doing this, we confirm the
results of Cardoso et al.~\cite{Cardoso} for their models, but also
show that the conclusions drawn were excessively restrictive. In particular we show that not all rotating gravastars are
unstable to the ergoregion instability. Rather, we find that
models of rotating gravastars without an ergoregion (and therefore
stable) can be constructed also for extreme rotation rates, namely for
models with $J/M^2 \ge 1$, where $J$ and $M$ are the gravastar's
angular momentum and mass, respectively.

The paper is organized as follows: in Section~\ref{sec:Model} we
briefly review our gravastar model and the slow-rotation
approximation. In Section~\ref{sec:Perturb} we develop the equations
for scalar perturbations and present the WKB approximation used. In
Section~\ref{sec:Discussion} we present our results and analyze the
behavior of the instability in the space of parameters and in
Section~\ref{sec:Conclusions} we present our concluding remarks. We
use $c = G = 1$ throughout the paper.

\section{A rotating gravastar model}
\label{sec:Model}

We start our analysis by adding a uniform rotation to the simple fluid
gravastar model with anisotropic pressure presented in
ref.~\cite{Chirenti}. We recall that the use of anisotropic pressures
was introduced by Cattoen et al.~\cite{Cattoen} to remove in part the
complications produced by the infinitesimal shells in the original
gravastar model of Mazur and Mottola~\cite{Mazur}. In this way, the
anisotropic pressure replaces the surface tension introduced by the
matching of the metric in the infinitesimally thin shells. Although
the use of an anisotropic pressure is essentially arbitrary, as
arbitrary are the equations of state used to describe such a pressure,
it has the appealing property of being continuous and thus of allowing
one to build equilibrium models without the presence of
infinitesimally thin shells and thus look more seriously into the
issue of stability.

We write therefore the line element for a finite-thickness rotating
gravastar with anisotropic pressures within the slow-rotation
approximation, (namely at first-order in the angular velocity in the
small parameter $\Omega/\Omega_K$, where $\Omega_K$ is the Keplerian
limit) as
\begin{eqnarray}
ds^2 = -e^{\nu(r)}dt^2 + e^{\lambda(r)}dr^2 +
r^2d\theta^2 + \nonumber \\
+ r^2\sin^2\theta(d\phi - \omega(r)dt)^2\,,
\label{metric}
\end{eqnarray}
together with the energy momentum tensor given by
\begin{equation}
T^{\mu}_{\phantom{\mu}\nu} = (\rho+p_{\rm t})u^{\mu}u_{\nu} + p_{\rm
  t}\delta^{\mu}_{\phantom{\mu}\nu} + (p_{\rm r}-p_{\rm t})s^{\mu}s_{\nu}\,,
\end{equation}
where $\rho$ is the energy density of the gravastar, $p_{\rm r}$ and
$p_{\rm t}$ are the radial and tangential pressures, respectively. The
vector $u_{\mu}$ is the fluid four-velocity,
\begin{eqnarray}
u^{\mu}u_{\mu} = -1\,, \quad u^r = u^{\theta} = 0\,, \quad u^{\phi}\,
= \Omega u^t\,, \nonumber \\
u^t = \left[ -\left( g_{tt} + 2\Omega g_{t\phi} + \Omega^2g_{\phi\phi}
  \right) \right]^{-1/2}\,, \nonumber
\end{eqnarray}
and the vector $s^{\mu}$ is orthogonal to the fluid four-velocity,
\begin{equation}
s^{\mu}s_{\mu} = 1\,, \quad u^{\mu}s_{\mu} = 0 \quad (s_t = s_{\phi} =
0)\,.
\end{equation}

The functions $\nu(r)$ and $\lambda(r)$ in the line element
(\ref{metric}) are given in terms of
the mass $m(r)$ and the radial pressure $p_r(r)$, as in the non-rotating
case,
\begin{eqnarray}
  \label{lambda}
  e^{-\lambda} = 1-\frac{2m(r)}{r}\,,\\
  \label{nu'}
  \nu ' = \frac{2m(r)+8\pi r^3p_r}{r(r-2m(r))}\,,
\end{eqnarray}
where a prime denotes a (total) radial derivative. As usual, we
define the gravitational mass within a radius $r$ as $m(r) \equiv \int_0^r
4\pi r^2\rho dr$ and
\begin{equation}
\rho(r) = \left\{ \begin{array}{ll}
    \rho_0\,, & 0 \le r \le r_1\\
    ar^3+br^2+cr+d\,, & r_1 < r < r_2\\
    0\,, & r_2 \le r
  \end{array} \right.\,,
\label{rho}
\end{equation}
with the coefficients $a,b,c,d$ given by
\begin{eqnarray}
  a = \frac{2\rho_0}{\delta^3}\,,\\
  b = -\frac{3\rho_0(r_2+r_1)}{\delta^3}\,,\\
  c = \frac{6\rho_0r_1r_2}{\delta^3}\,,\\
  d = \frac{\rho_0(r_2^3-3r_1r_2^2)}{\delta^3}\,,
\end{eqnarray}
where $\delta \equiv r_2-r_1$ is the ``thickness'' of the gravastar and 
\begin{equation}
\rho_0 = \frac{15M}{2\pi(r_1+r_2)(2r_1^2+r_1r_2+2r_2^2)}\,.
\end{equation}

The equation of state used for $p_r(\rho)$ serves here only as a
closure relation and is therefore chosen to be in the simplest
possible form, namely a polynomial of the
type~\cite{Mbonye,DeBenedictis}
\begin{equation}
  p_r(\rho) =
  \left[\alpha-(\alpha+1)\left(\frac{\rho}{\rho_0}\right)^2\right] 
  \left(\frac{\rho}{\rho_0}\right)\rho\,,
  \label{eos}
\end{equation}
where $\alpha = 2.2135$ is determined by demanding that the maximum
sound speed $c^2_{\rm s}$ at which ${d^2 p_r}/{d\rho^2} = 0$ coincides
with the speed of light to rule out a superluminal behavior. Finally,
the tangential pressure $p_t$ is given by the anisotropic TOV equation
\begin{equation}
  p_t = p_r+\frac{r}{2}p_r'+\frac{1}{2}(p_r+\rho)
  \left[\frac{m(r)+4\pi r^3p_r}{r(1-2m(r)/r)}\right]\,.
  \label{p_t}
\end{equation}

The function $\omega(r)$ can be shown to be of first order in the
angular velocity $\Omega$ and it describes the dragging of inertial
frames. Formally, it can be obtained from the $^t_{\ \ \phi}$
component of the field equations~\cite{Hartle}  
\begin{equation}
  R_{\phi}^{\phantom{\phi}t} = 8\pi T_{\phi}^{\phantom{\phi}t}\,,
\end{equation} 
which gives, for our anisotropic case,
\begin{equation}
\varpi'' + \left(\frac{4}{r} - \frac{4\pi r^2(\rho+p_r)}{r-2m}\right)\varpi'
= \frac{16 \pi r(\rho+p_t)}{r-2m}\varpi\,,
\label{dragging}
\end{equation}
where 
\begin{equation}
\varpi(r) = \Omega - \omega(r)\,,
\label{def_varpi} 
\end{equation}
for the region of the spacetime interior to the gravastar. It
is easy to see that eq.~(\ref{dragging}) is equivalent to eq.~(2.25)
of ref.~\cite{Cardoso}. In the
exterior region, on the other hand, we have
\begin{equation}
  \varpi(r) = \Omega - \frac{2J}{r^3}\,, \quad \textrm{or} \quad
  \omega(r) = \frac{2J}{r^3}\,.
\label{varpi_ext}
\end{equation}
The definition of $\varpi$ in eq.~(\ref{def_varpi}) deserves some
attention and it corresponds to the difference between $\Omega$, the
angular velocity of the gravastar (as seen by an observer at rest),
and $\omega(r)$, which gives the angular velocity of a zero angular
momentum observer (ZAMO). Therefore, $\varpi(r)$ is the angular
velocity of the gravastar as seen by the ZAMO.

As a general result in the asymptotically flat limit, the dragging
must go to zero as $\omega = 2J/r^3 + O(r^{-4})$ for $r \to \infty$,
thus defining $J$ as the angular momentum of the
spacetime~\cite{Bardeen}. Demanding now that the exterior equations
(\ref{varpi_ext}) to be consistent with this asymptotic limit implies
immediately that the integration constant $J$ in (\ref{varpi_ext})
should be identified as the total angular momentum of the gravastar.

\begin{figure}[tp]
\begin{center}
  \includegraphics[angle=270,width=1.0\linewidth]{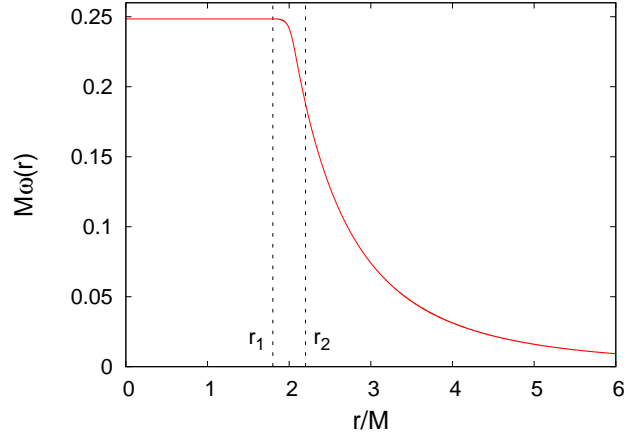}
\end{center}
  \caption{Typical example for the dragging of inertial frames
    $\omega(r)$, for a gravastar with $\mu = 0.45$, $\delta/M = 0.4$
    and $J/M^2 = 1$ ($\Omega/\Omega_K = 0.82$). We can see that
    $\omega = {\rm const.}$ in the interior ($r < r_1$) and $\omega
    \to 0$ in the exterior region ($r > r_2$).}
  \label{fig_drag}
\end{figure}

A representative behavior of the frame-dragging function $\omega(r)$
is shown in Fig.~\ref{fig_drag}, for a gravastar with thickness of the
shell $\delta/M = 0.4$ and compactness $\mu \equiv M/r_2 = 0.45$. 
The frame-dragging is obtained by
numerically integrating eq.~(\ref{dragging}), with initial conditions
$\varpi'(r=0) = 0$ and $\varpi(r=0)$ finite. As shown in the figure, the
solution for $\omega(r)$ in the interior of the gravastar is constant
[\cf eqs. (\ref{dragging}) and (\ref{def_varpi})] and monotonically
decreases outwards. The two integration constants $\Omega$ and $J$ can
then be determined by matching the interior and exterior solutions at
the boundary $r = r_2$. Finally, in the exterior of the gravastar,
$\omega(r)$ goes asymptotically to zero [\cf eq.~(\ref{varpi_ext})].

\section{Scalar perturbations and the ergoregion instability}
\label{sec:Perturb}

We recall that if a compact relativistic star rotates sufficiently
fast, it will posses an ergoregion, namely a region in which the
frame-dragging is so intense that stationary orbits are no longer
possible. Within this region even massless particles, such as photons,
will be involved in the ``drag'' and all trajectories will rotate in
the prograde direction. Mathematically, the boundary of the ergoregion
is defined as where the covariant $tt$-component of the metric has a zero, \ie
\begin{equation}
g_{tt} = -e^{\nu} + r^2 \omega^2 \sin^2 \theta = 0\,,
\label{ergo}
\end{equation}
and it should be noted that the ergoregion does not need to be
restricted to the exterior of the compact star and, rather, it can
also involve regions interior to the star.

The main interest in the existence of an ergoregion in a compact star
stems from the fact that these regions lead to a secular instability
by means of which any initially small perturbation will grow
exponentially in time~\cite{Comins, Cardoso}. Clearly, one expects
that nonlinear effects will intervene to limit the growth of the
instability once this has reached a sufficiently high (saturation)
amplitude which, however, cannot be determined on the basis of a
linear perturbative investigation. While determining such an amplitude
is of great physical interest, we will here limit ourselves to a
simple linear analysis of the problem. On the other hand, we will
determine not only if an instability can or cannot develop but also,
and more importantly, the characteristic timescale for the growth of
the instability. Indeed, as shown already many years ago in
ref.~\cite{Comins}, for ultra-compact uniform-density stars, the
timescale for the growth of scalar perturbations via the ergoregion
instability can sometimes be several orders of magnitude larger than
the age of the universe. When this is case, the instability grows so
slowly that the stars are effectively stable even if the linear
analysis reveals that they are mathematically unstable.

Hereafter we will concentrate on scalar perturbations and within the
WKB approximation. There are many references in the literature where
this method is described, but we suggest for instance
ref.~\cite{Simmonds}. The WKB method (from Wentzel, Kramers,
Brillouin) is used to approximate the solution of differential
equations of the general form
\begin{equation}
y''(x) +  \omega^2r(x)y(x) = 0\,,
\label{generic_WKB}
\end{equation}
where $\omega^2 \gg 1$. As we will see below, the scalar perturbations
that we are considering are described by eq.~(\ref{scalar}), which has
the same form as eq.~(\ref{generic_WKB}) above. The same equation is
often found to arise in quantum mechanics, but also in classical physics
problems. The key point of the approximation is the assumption
$\omega^2 \gg 1$ (high frequency approximation in quantum
mechanics). It allows us to consider that, for every $\omega_n \gg 1$,
then $r(x)$ is approximately constant between two zeros of
$y_n(x)$. It is easy to see that this approximation greatly simplifies
the integration of eq.~(\ref{generic_WKB}). 

Of course, studying the response of a rotating
gravastar to electromagnetic or gravitational perturbations would be
astrophysically more interesting and realistic. However, the use of
scalar perturbations has the important advantage that in this case one
can decompose the perturbations in spherical harmonics and reduce the
perturbation equation to a single ordinary differential equation,
which can be integrated with very modest computational costs and high
accuracy. In addition, because the order of magnitude for the growth
of the perturbations is expected to be the same for scalar,
electromagnetic and gravitational perturbations [this is indeed the
  case for the decay of perturbations of different spins in a
  Schwarzschild spacetime (see ref.~\cite{Konoplya} for some numerical
  results)], estimating the growth time of scalar perturbations allows
for a simple and direct extension also to other types of
perturbations.

Bearing this in mind, we next proceed to the study of the massless
scalar wave equation in the slowly rotating gravastar background that
we introduced in Section~\ref{sec:Model}. We briefly review here the
basic steps usually followed in this type of analysis; such steps can
be found in many papers and are reproduced here only for
completeness. The wave equation
\begin{equation}
\frac{1}{\sqrt{-g}}\frac{\partial}{\partial x^{\mu}} \left(
\sqrt{-g}g^{\mu \nu} \frac{\partial \psi}{\partial x^{\nu}} \right) =
0\,,
\label{KG_eq}
\end{equation}
can be separated by using the \textit{ansatz}
\begin{eqnarray}
\psi(t,r,\theta,\phi) &=& \bar{\chi}(r) \exp \left\{ -\frac{1}{2} \int
  \left( \frac{2}{r} + \frac{\nu'}{2} - \frac{\lambda'}{2} \right)dr
  \right\} \times \nonumber \\
&\times& e^{{\rm i}\sigma t} Y_{\ell m}(\theta,\phi)\,. 
\label{def_chi}
\end{eqnarray}
where the eigenfrequency $\sigma$ is in general complex, with $\sigma
= \sigma_r - i/\tau$. As a result, $\tau > 0$ leads to an
exponential growth, while $\tau < 0$ leads to a decay.

In the high $m$ limit (and taking $\ell = m$), eq.~(\ref{KG_eq}) can
be written using eq.~(\ref{def_chi}) and keeping only the dominant
terms (\ie first order in $\omega$ and $1/m$) as
\begin{equation}
\bar{\chi}_{,rr} + m^2T(r,\Sigma)\bar{\chi} = 0\,,
\label{scalar}
\end{equation}
where $m$ is here the order of the $Y_{\ell m}(\theta,\phi)$ spherical
harmonics (not to be confused with the mass function $m(r)$), $\Sigma
= \sigma/m$ is the negative of the pattern speed of the perturbation
and $T$ is given in terms of the two rotationally split ``effective
potentials'' $V_+$ and $V_-$,
\begin{eqnarray}
\label{def_T}
  T & \equiv & e^{\lambda-\nu}(\Sigma - V_+)(\Sigma - V_-)\,,\\
\label{def_Vpm}
  V_{\pm} & \equiv & -\omega \pm \frac{e^{\nu/2}}{r}\,.
\end{eqnarray}
It is easy to see that the boundary of the ergoregion given in
eq.~(\ref{ergo}) coincides, in the equatorial plane, with $V_+ = 0$,
so that the ergoregion is effectively contained in the region where
$V_+ < 0$.

\begin{figure}[htp!]
\begin{center}
  \includegraphics[angle=270,width=1.0\linewidth]{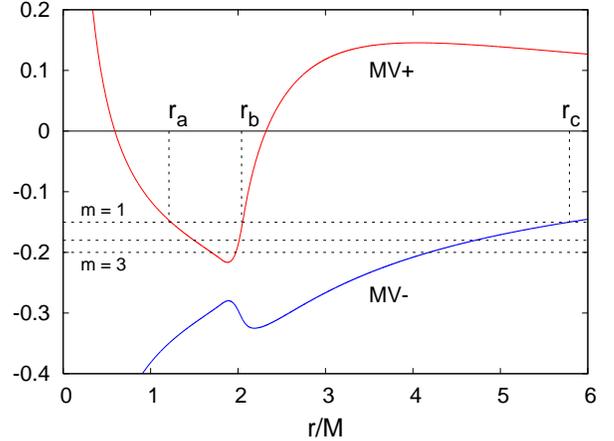}
\end{center}
  \caption{Typical example for the potentials $V_{\pm}$, for a
  gravastar with $\mu = 0.45$, $\delta/M = 0.4$ and $J/M^2 = 1$
  ($\Omega/\Omega_K = 0.82$). The first three unstable modes with
  negative energy ``trapped'' in the potential well are depicted, as
  well as the points $r_a$, $r_b$ and $r_c$ for the $\ell=m=1$ mode.}
  \label{fig_well}
\end{figure} 

Our stability analysis translates therefore to finding the complex
eigenfrequencies $\sigma$ of the scalar wave modes that satisfy
eq.~(\ref{scalar}). However, because these frequencies generally have
the real part being much larger than the imaginary one, \ie,
$\Re(\sigma) \gg \Im(\sigma)$ (or, equivalently, $\sigma_r \gg
1/\tau$), it is a reasonable approximation to consider $\sigma$ as
essentially a real number. It is also more
convenient to the problem to use $\Sigma = \sigma/m$ instead of
$\sigma$ itself (see the Appendix for more details). 

\begin{figure*}[htp!]
\begin{center}
  \includegraphics[angle=270,width=0.49\linewidth]{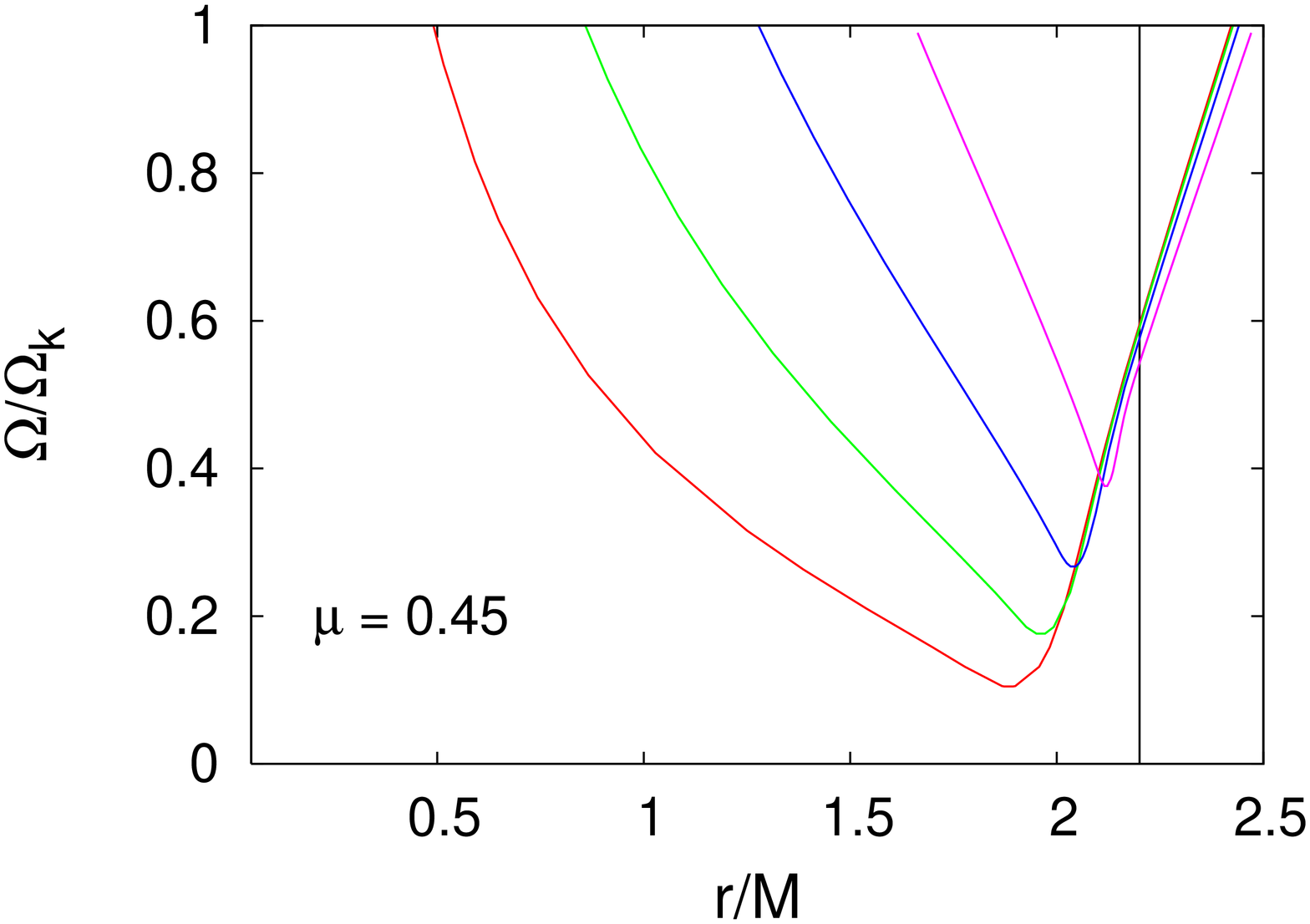}
  \hskip 0.2cm
  \includegraphics[angle=270,width=0.49\linewidth]{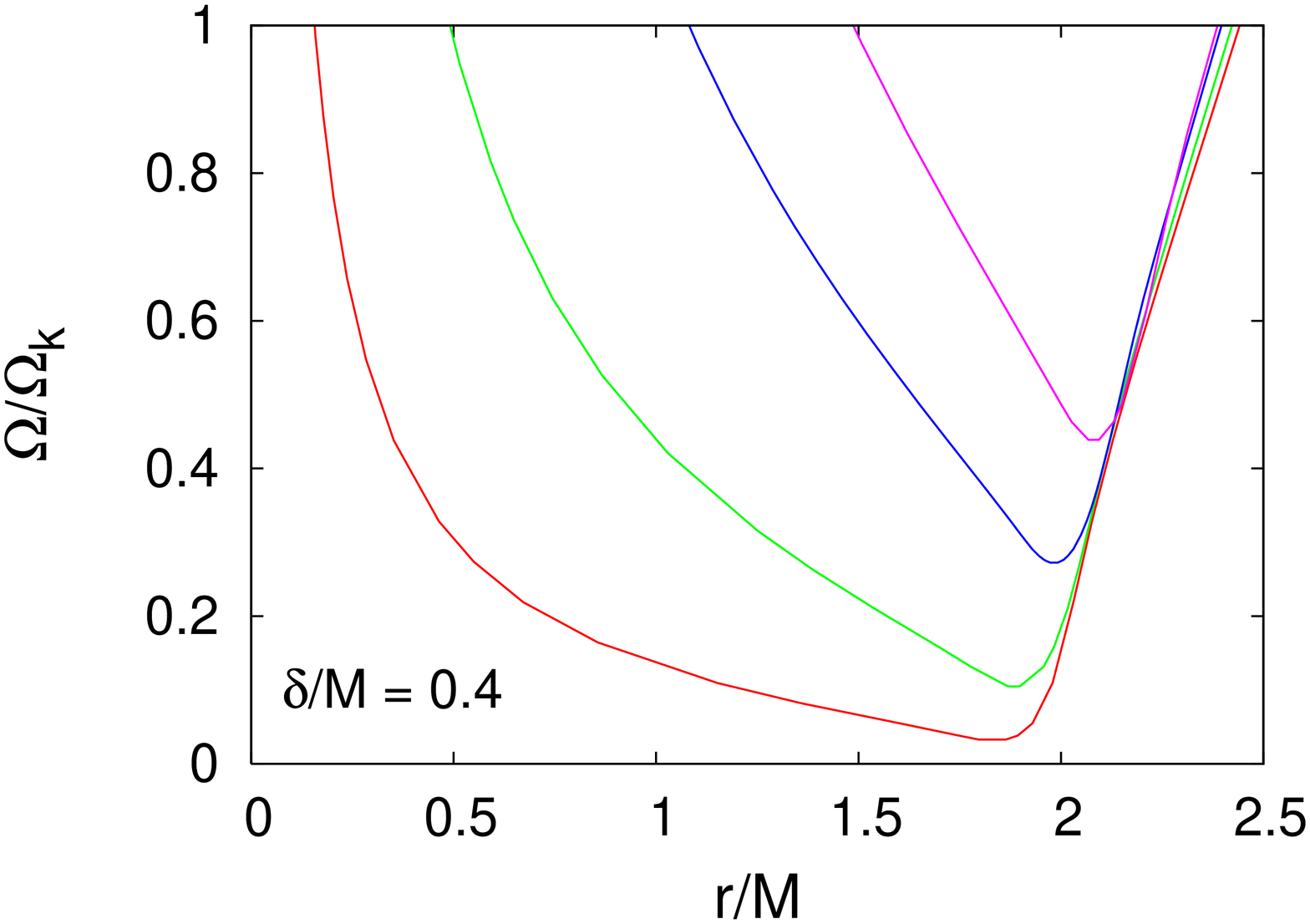}
\end{center}
  \caption{\textit{Left panel:} Change in the size of the ergoregion
    for fixed compactness $\mu = 0.45$, several different values for
    $\delta/M$ (from top to bottom $\delta/M = 0.1,\,0.2,\,0.3\,$ and
    0.4) and increasing angular velocity $\Omega$, until the Keplerian
    limit $\Omega_K$. A given value of $\Omega$ on the vertical axis
    determines the inner and outer radii of the ergoregion, while the
    vertical line shows the location of the radius of the gravastar
    for all the models: the ergoregion generally starts in the
    interior of the gravastar and goes up to a radius exterior to the
    radius $r_2$ of the gravastar. \textit{Right panel:} Same as the
    left panel, but for fixed thickness of the shell $\delta/M = 0.4$
    and different values for $\mu$ (from top to bottom $\mu =
    0.42,\,0.43,\,0.45$ and 0.47).  In this case we do not show the
    surface radius, since each model has a different radius $r_2$.}
  \label{region_mu}
\end{figure*} 

Furthermore, when considered within the WKB approximation, the
potentials $V_{\pm}(r)$ determine four different regions which are
reminiscent of those appearing for waves trapped in the potential well
of $V_+$. To illustrate this in more detail we show in
Fig.~\ref{fig_well} an example of the two potentials for some typical
parameters. In addition, the figure shows with dotted horizontal lines
the frequencies of the first three unstable modes, \ie $\Sigma_{m=1},
\Sigma_{m=3}$, so that $\Sigma^2$ can be viewed as an analog of the
energy of a quantum mechanical particle.  For each of the unstable
modes there is therefore an inner ``forbidden'' region $0<r<r_a$, an
``allowed'' region $r_a<r<r_b$, a ``potential barrier'' $r_b<r<r_c$
and an external ``allowed'' region $r>r_c$. The points $r_a$, $r_b$
and $r_c$ shown in the figure correspond to the $\ell=m=1$ mode.

The WKB matching of the wave functions in the four regions is shown in
the Appendix, where we present the derivation of eqs.~(\ref{BS}) and
(\ref{tau}) below. The unstable modes are determined by requiring that
\begin{equation}
  m\int_{r_a}^{r_b}\sqrt{T}dr = \left(n+\frac{1}{2}\right)\pi\,;\quad
  n = 0,1,2,\ldots
\label{BS}
\end{equation}
and
\begin{equation}
  \tau = 4 \exp \left( 2m \int_{r_b}^{r_c} \sqrt{|T|}dr \right)
  \int_{r_a}^{r_b}\frac{d}{d\Sigma}\sqrt{T}dr\,,  
\label{tau}
\end{equation}
where
\begin{equation}
  \frac{d}{d\Sigma}\sqrt{T} = (\Sigma + \omega)\frac{e^{\lambda -
  \nu}}{\sqrt{T}}\,. 
\label{dTdSigma}
\end{equation}
Equation~(\ref{BS}) is the classical Bohr-Sommerfeld rule and
determines $\Sigma$, while eq.~(\ref{tau}) gives the growing time of
the instability. The limits of
the integration interval $r_a$, $r_b$ and $r_c$ have the physical
interpretation given above for Fig.~\ref{fig_well} and correspond,
mathematically, to the turning points as given by the condition $V_+ =
\Sigma$ (or $T = 0$) and to the beginning of the free allowed region
as given by the condition $V_- = \Sigma$. Note also that the use of
the absolute value for $T$ in the first integral of eq.~(\ref{tau}) is
due to the fact that in the interval $r_b<r<r_c$ (\ie inside the
potential barrier) $T<0$ [\textit{cf.} eq.~(\ref{def_T})].

It is customary in the evaluation of the integral in eq.~(\ref{BS}),
to use an analytical parabolic approximation for $V_+$ (see
ref.~\cite{Comins}). When this is done, the functions $\lambda$, $\nu$
and $V_-$ are taken to be constants, with their values set to
$\lambda(R)$, $\nu(R)$ and $V_-(R)$, where $R$ is the radius at which
$V_+$ has its minimum. This greatly simplifies the calculations and is
a very useful strategy in general. In the case of a gravastar,
however, the potential well in $V_+$ is typically very asymmetric
around the minimum of the potential, so the usual analytical parabolic
approximation is not appropriated in this case, for it would introduce
too large errors and it has not been used in the numerical solution of
eq.~(\ref{tau}). 

Collected in Table~\ref{table1} are some typical numerical values
obtained for the potential $V_+$ of the gravastar with $\mu = 0.45$
and $\delta/M = 0.4$ and different values of $J$. With these data it
is possible to confirm that both the size of the ergoregion (which
extends from $r_{\rm min}$ , the inner boundary of the ergoregion, to
$r_{\rm max}$, the outer boundary of the ergoregion, in the notation of
the table) and the depth of the potential well $|V_+(R)|$ increase
with $J$, while $R$ remains essentially unchanged.

\begin{table}
\caption{\label{table1}Typical values of $J/M^2$, $\Omega/\Omega_K$,
  $r_{\rm min}$ (inner boundary of the ergoregion), $r_{\rm max}$
  (outer boundary of the ergoregion), $R$ (radius at which $V_+$ is
  minimum) and $V_+(R)$ for the gravastar with $\mu = 0.45$ and
  $\delta/M = 0.4$.}
\begin{ruledtabular}
\begin{tabular}{llllll}
$J/M^2$ & $\Omega/\Omega_K$ & $r_{\rm min}/M$ & $r_{\rm max}/M$ &
$R/M$ & $MV_+(R)$\\ 
\hline
1.2 & 0.98 & 0.500 & 2.411 & 1.8789 & -0.2661\\
1.0 & 0.82 & 0.591 & 2.321 & 1.8797 & -0.2164\\
0.8 & 0.65 & 0.721 & 2.231 & 1.8806 & -0.1668\\
0.6 & 0.49 & 0.917 & 2.146 & 1.8815 & -0.1172\\
0.4 & 0.33 & 1.225 & 2.071 & 1.8824 & -0.0068\\
0.2 & 0.16 & 1.683 & 1.987 & 1.8834 & -0.0018\\
\end{tabular}
\end{ruledtabular}
\end{table}

\section{Results}
\label{sec:Discussion}

We have first checked our results against the values obtained by
Cardoso {\it et al} in ref.~\cite{Cardoso} (for $\Sigma$ and $\tau$)
for a typical model with $r_1 = 1.8$, $r_2 = 2.2$ and $M = 1$, with
very good agreement. Based on their findings for this typical model,
which shows a very rapidly growing instability, they have concluded
that the ergoregion instability would ``rule out'' the possibility of
rotating gravastars to exist as alternatives to Kerr black
holes. However, drawing a general conclusion from a specific example
can be too restrictive. In view of this, we
have decided to reconsider the issue of whether all of the rotating
gravastars are unstable to the ergoregion instability and if so over
what timescales. As we will show in this Section, by considering a
much larger space of parameters we reach conclusions which are rather
different from those of ref.~\cite{Cardoso}.

We start by showing in Fig.~\ref{region_mu} how the size of the
ergoregion (taken at the equatorial plane) depends on the parameters
of the gravastar and in 
particular on its compactness $\mu$, on its thickness $\delta$ and on
its angular velocity $\Omega$, which we take to range from zero to the
Keplerian (mass shedding) limit $\Omega_K \equiv \left( {M}/{r_2^3}
\right)^{1/2}$.

On the left panel of Fig.~\ref{region_mu}, in particular, we report
the variation of the size of the ergoregion for models with $\mu =
0.45$ and different values of $\delta$, with the vertical line signing
the surface of the gravastar. On the right panel, on the other hand,
we obtain essentially the same behavior, but this time for models with
$\delta/M = 0.4$ and different values of $\mu$. These plots are to be
interpreted as follows: for a given value of $\Omega/\Omega_K$ on the
vertical axis, each of the curves in the plot provides the values of
the inner and outer boundaries of the ergoregion. It is then easy to
see that the ergoregion becomes obviously larger for increasing values
of $\Omega$, but also for increasing values of $\delta$ and $\mu$. It
also should be noted that the ergoregion is mostly (but not
exclusively) contained in the interior of the gravastar, and it also
extends to the exterior region.  Most importantly, however, the local
minima in Fig.~\ref{region_mu} indicate the first important result of
this investigation: \textit{not all rotating gravastars possess an
  ergoregion}. Rather, for any choice of compactness and thickness of
the gravastar, there exists a minimum angular velocity $\Omega_{\rm
  min}$ above which an ergoregion of finite size develops. In this
respect, and not surprisingly, gravastars behave like rotating compact
stars. It is then a trivial consequence of the above result that not
all rotating gravastars are unstable to the ergoregion instability.

\begin{figure}[tp!]
\begin{center}
  \includegraphics[angle=270,width=1.0\linewidth]{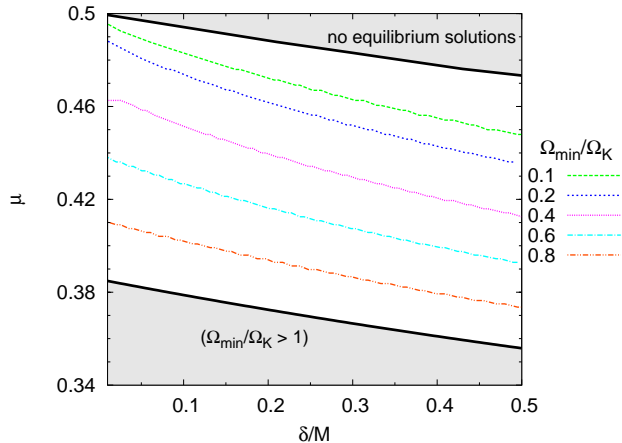}
\end{center}
  \caption{Minimum angular velocity necessary for the existence of an
    ergoregion, as a function of both $\mu$ and $\delta$. If an
    ergoregion is present, the instability will set in for a high
    enough value of $m$. In the grey area at the bottom, an angular
    velocity $\Omega$ larger than the mass shedding limit $\Omega_K$
    would be required for an ergoregion to develop. The grey area at
    the top shows a constraint on the possible (nonrotating) gravastar
    solutions found in ref.~\cite{Chirenti}.}
  \label{contour}
\end{figure} 

This conclusion is summarized in Fig.~\ref{contour}, which shows the
minimum angular velocity necessary for the existence of an ergoregion,
as a function of both the compactness $\mu$ and thickness $\delta$ of
the gravastar. In other words, for any couple of values of $(\mu,
\delta)$ outside of the shaded regions, the corresponding gravastar
will not possess an ergoregion if spinning below the value of
$\Omega_{\rm min}$ at that point. Note also that in the grey area at
the bottom of the figure, an angular velocity $\Omega$ larger than the
mass shedding limit $\Omega_K$ would be required for an ergoregion to
develop. Similarly, the grey area at the top of the figure shows the
constraints on the possible (nonrotating) gravastar solutions found in
ref.~\cite{Chirenti}. It is important to remark that, as clearly shown
in Fig.~\ref{contour}, the minimum angular velocity $\Omega_{\rm min}$
increases with decreasing compactness of the gravastars. This
behavior sets an additional constraint on the parameters of the
gravastars that will be subject to the ergoregion instability: less
compact gravastars would have to rotate with $\Omega > \Omega_K$ to
form an ergoregion and are therefore also free from the instability.

\begin{figure}[tp!]
\begin{center}
  \includegraphics[angle=270,width=1.0\linewidth]{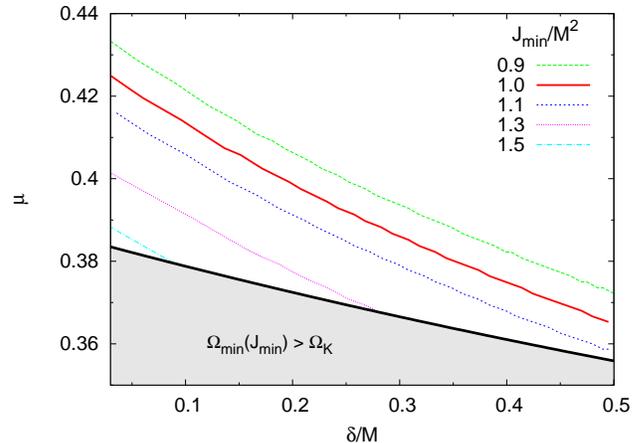}
\end{center}
  \caption{Minimum angular momentum necessary for the existence of an
  ergoregion, as a function of both $\mu$ and $\delta$. Gravastars
  with parameters $(\mu,\delta)$ given below a curve labeled with some
  value of $J_{\rm min}/M^2$ will be stable if rotating with angular momentum
  smaller or equal to $J_{\rm min}$. Note that the $\Omega$ is not constant
  along the lines, and increases in the direction of the region
  labeled $\Omega_{\rm min}(J_{\rm min}) > \Omega_K$.}
  \label{contour2_new}
\end{figure}

Interestingly, there is a non-small portion of the space of parameters
$(\mu, \delta)$ where very rapidly rotating gravastars exist, do not
possess an ergoregion and are therefore stable. We recall, in fact,
that while black holes have their angular momentum bounded by the Kerr
limit (\ie $J/M^2 \le 1$), stars (and gravastars!)  are not subject to
this constraint. As a result, as long as they are spinning below the
mass-shedding limit, gravastar models can be built that are stable and
even have $J/M^2 > 1$. This is shown in Fig.~\ref{contour2_new}, which
reports the minimum angular momentum necessary for the existence of an
ergoregion, as a function of both $\mu$ and $\delta$. Gravastars with
parameters $(\mu, \delta)$ given below a curve labeled with some value
of $J_{\rm min}/M^2$ will be stable if rotating with angular momentum
smaller than $J_{\rm min}$ (models on the curve are marginally stable,
\ie with $\tau = \infty$, for that value of $J_{\rm min}$). This is
the second important result of this paper: \textit{not all
  ultra-compact astrophysical objects rotating with $J/M^2 \sim 1$
  must be black holes}. This conclusion is thus less restrictive than
the one drawn by Cardoso et al.~ in ref.~\cite{Cardoso}.

\begin{figure}[tp!]
\begin{center}
  \includegraphics[angle=270,width=1.0\linewidth]{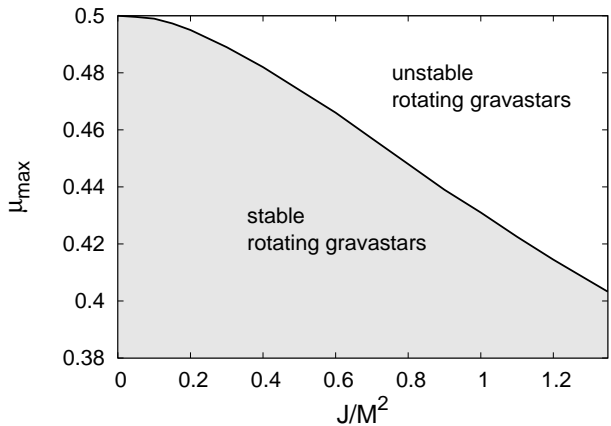}
\end{center}
  \caption{Maximum compactness $\mu_{\rm max}$ that a stable gravastar
    can have in terms of the angular momentum $J/M^2$. For a given
    value of $J$, gravastars with $\mu > \mu_{\rm max}(J)$ will be
    unstable. The maximum value for the compactness is obtained for
    $\delta \to 0$. This figure should be compared with the
    corresponding Fig.~1 of ref.~\cite{Chirenti}.}
  \label{mu_max}
\end{figure}

Note that the $\Omega$ is not constant along the $J_{\rm
  min}-$const. lines but, rather, it increases in the direction of the
region labeled $\Omega_{\rm min}(J_{\rm min}) > \Omega_K$. However, as
the angular momentum is increased, the maximum compactness allowed for
the stable models is also reduced, thus indicating a new bound on the
compactness of stable models. We show this additional constraint on
the gravastar's compactness in Fig.~\ref{mu_max}, which reports the
maximum allowed compactness $\mu_{\rm max}$ as a function of the
angular momentum $J/M^2$. [This figure should be compared with the
  corresponding Fig.~1 of ref.~\cite{Chirenti} where the constraint on
  the compactness was shown as a function of the thickness for
  nonrotating gravastars.]

Focusing on $J/M^2 \simeq 1$, a case which is astrophysically very
interesting~\cite{McClintock,Rezzollaetal03}, Fig.~\ref{mu_max} shows
that it is possible to construct stable rotating gravastars with
$J/M^2 \simeq 1$ as long as their compactness is less than $\mu_{\rm
  max} \lesssim 0.43$. Such a compactness is clearly smaller than that
of a black hole with the same angular momentum (\ie $\mu=1$), but much
larger than the typical compactness for compact stars
and neutron stars, \ie $0.15 \lesssim \mu \lesssim 0.2$.  (We recall
that the maximum compactness for for perfect fluid nonrotating spheres
is given by the Buchdahl-Bondi limit $\mu = 4/9 \simeq
0.44$~\cite{Buchdahl1,Martin}). It is still unclear whether
astronomical electromagnetic observations will (ever) be able to
distinguish a stable rapidly rotating gravastar with $\mu \sim 0.43$
from a rotating black hole with $J/M^2\sim 1$ (see the discussion in
ref.~\cite{AKL02}). Yet, the constraints emerging from
Fig.~\ref{mu_max} provide other means (besides the measurement of
quasi-normal modes~\cite{Chirenti}) in which astronomical observations
could be used 
to distinguish (rotating) gravastars from (rotating) black holes.

So far all of the considerations made were on the general properties
of rotating gravastars. However, to fix the ideas and also provide
some reference numbers on the frequencies and timescales for the
ergoregion, we now discuss in more detail one typical case which we
will assume to be our reference model (\cf Table~\ref{table1}). More
specifically, we consider a gravastar with mass $M = 10M_{\astrosun}$
and inner and outer radii $r_1 = 1.8 M \simeq 27\,{\rm km}$, $r_2 =
2.2M \simeq 32\,{\rm km}$ (or, alternatively, $\mu \simeq 0.45,
\delta/M = 0.4$). Its mass-shedding spin frequency will be $\nu_K =
\Omega_K/2\pi \simeq 990\,{\rm Hz}$. Such a gravastar will be stable
(because without ergoregion), for $J/M^2 \lesssim 0.13$ (or
$\Omega/\Omega_K \lesssim 0.10$). Conversely, if the gravastar is set
to rotate at a higher rate, namely with $J/M^2 \simeq 0.22$ (or
$\Omega/\Omega_K \simeq 0.18$), an ergoregion will be present and the
instability will develop over a timescale $\tau \sim 10^{17}\,{\rm
  s}$, with $\ell=m=4$ being the lowest unstable mode. Note that such
a timescale is comparable with the Hubble timescale and thus the
rotating gravastar will be physically stable although mathematically
unstable (a similar result holds true also for compact uniform-density
stars~\cite{Comins}). In practice, it is necessary to spin the same
gravastar up to $J/M^2 \simeq 0.61$ (or $\Omega/\Omega_K \simeq 0.50$)
in order for the instability to develop on a timescale of the order of
$1\ {\rm s}$, with the $\ell=m=1$ being the lowest unstable mode.

\begin{figure*}[htp!]
  \includegraphics[angle=270,width=0.49\linewidth]{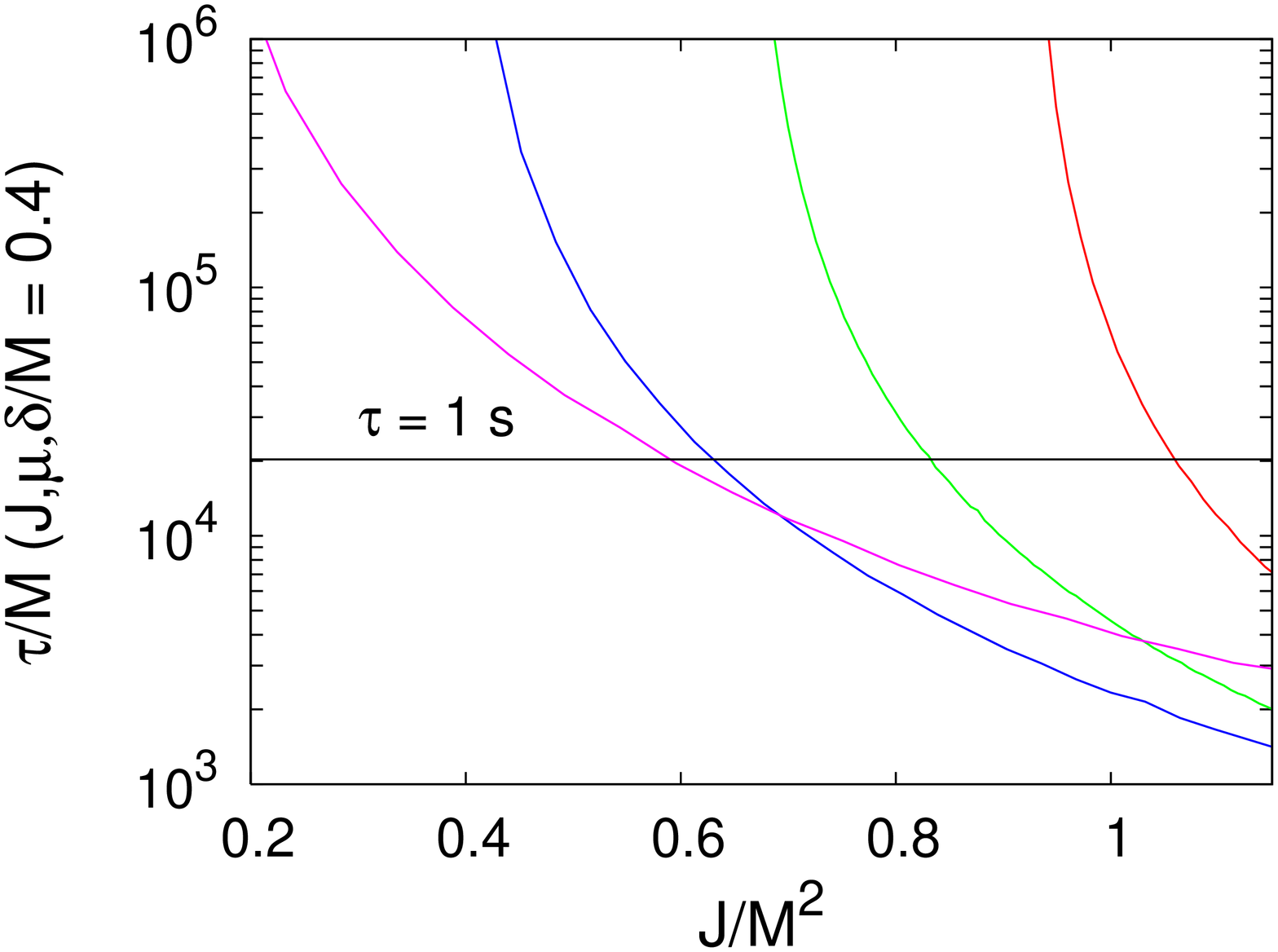}
  \hskip 0.2cm
  \includegraphics[angle=270,width=0.49\linewidth]{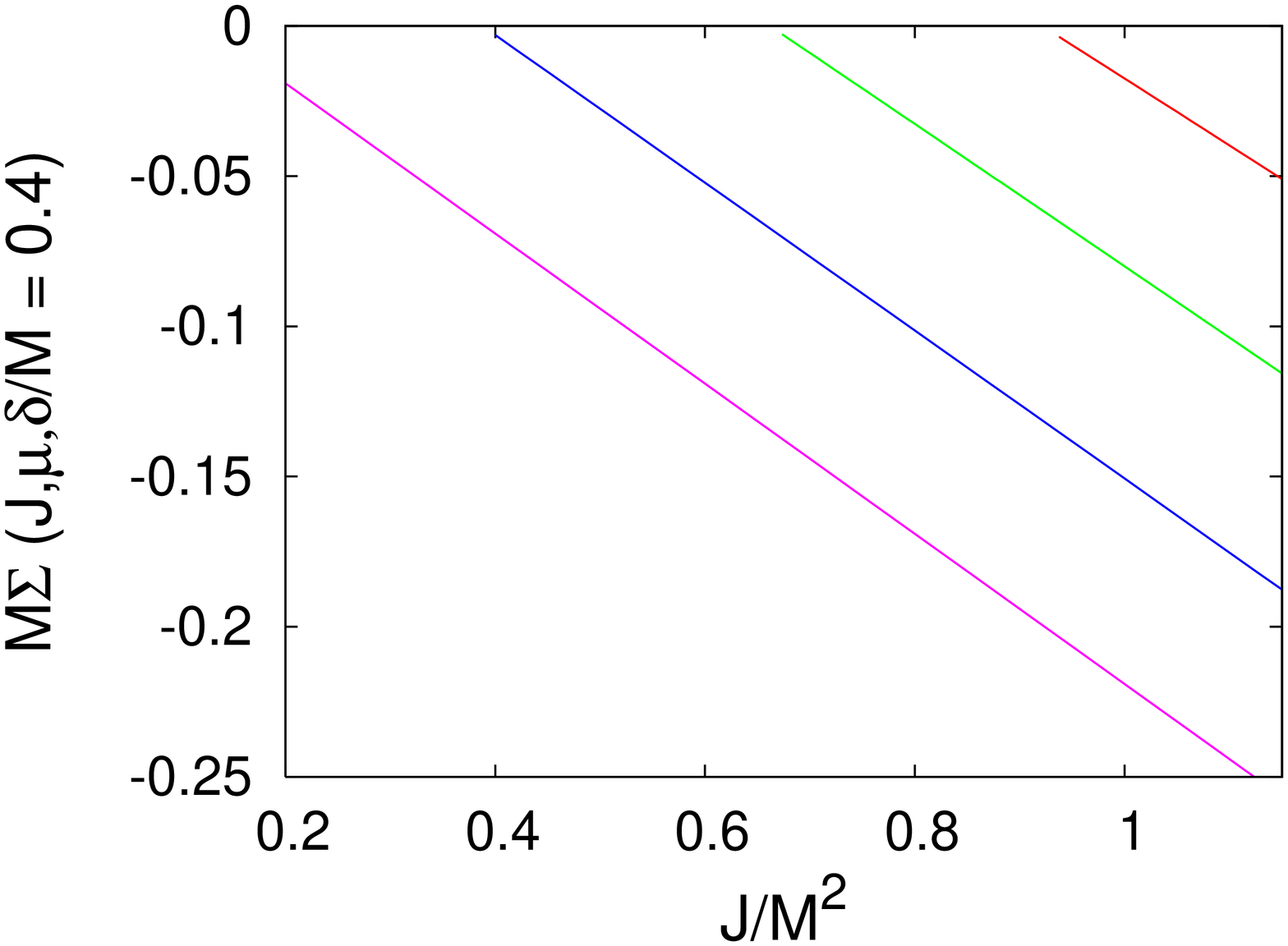}
  \vskip 0.2cm
  \includegraphics[angle=270,width=0.49\linewidth]{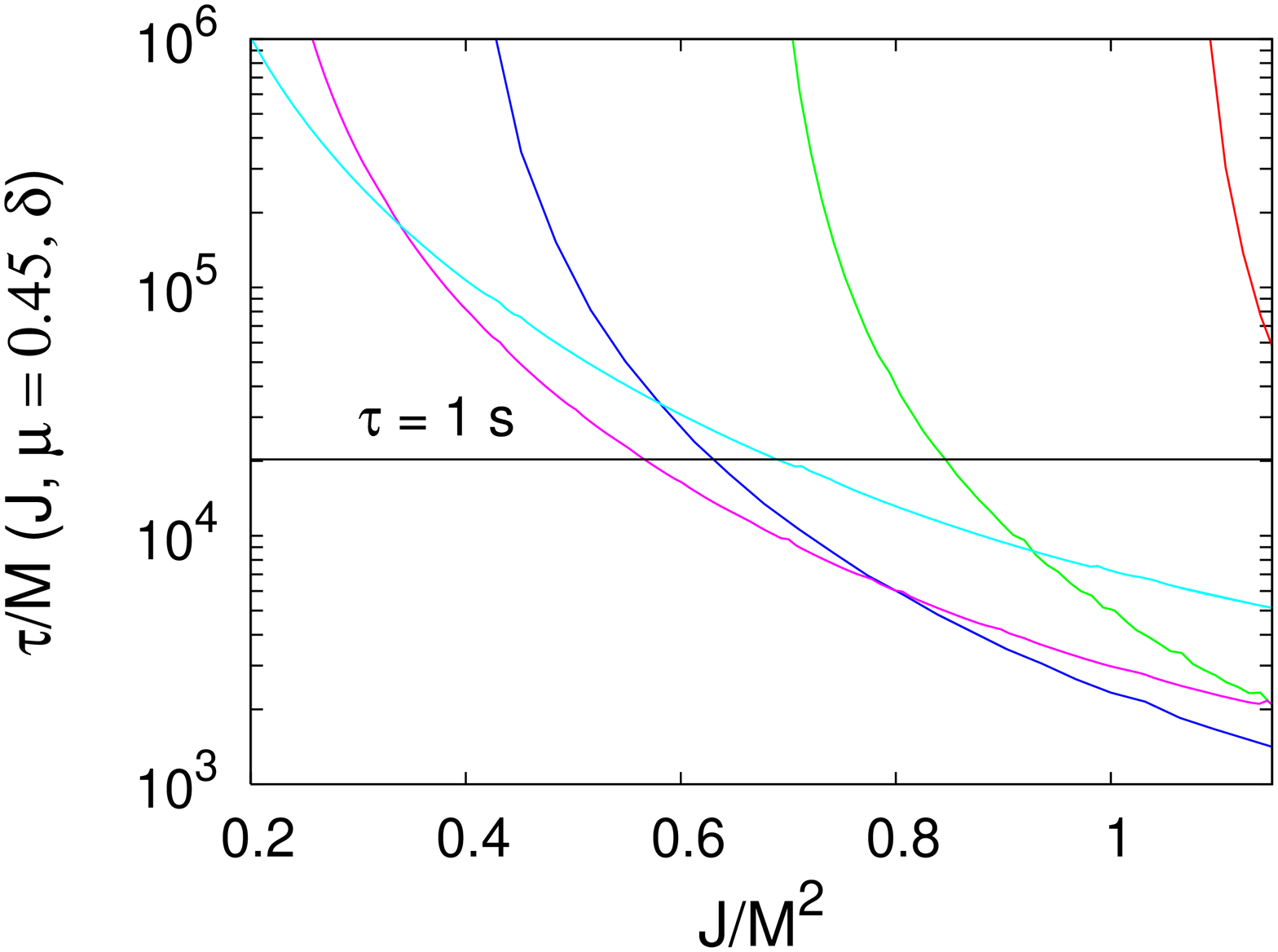}
  \hskip 0.2cm
  \includegraphics[angle=270,width=0.48\linewidth]{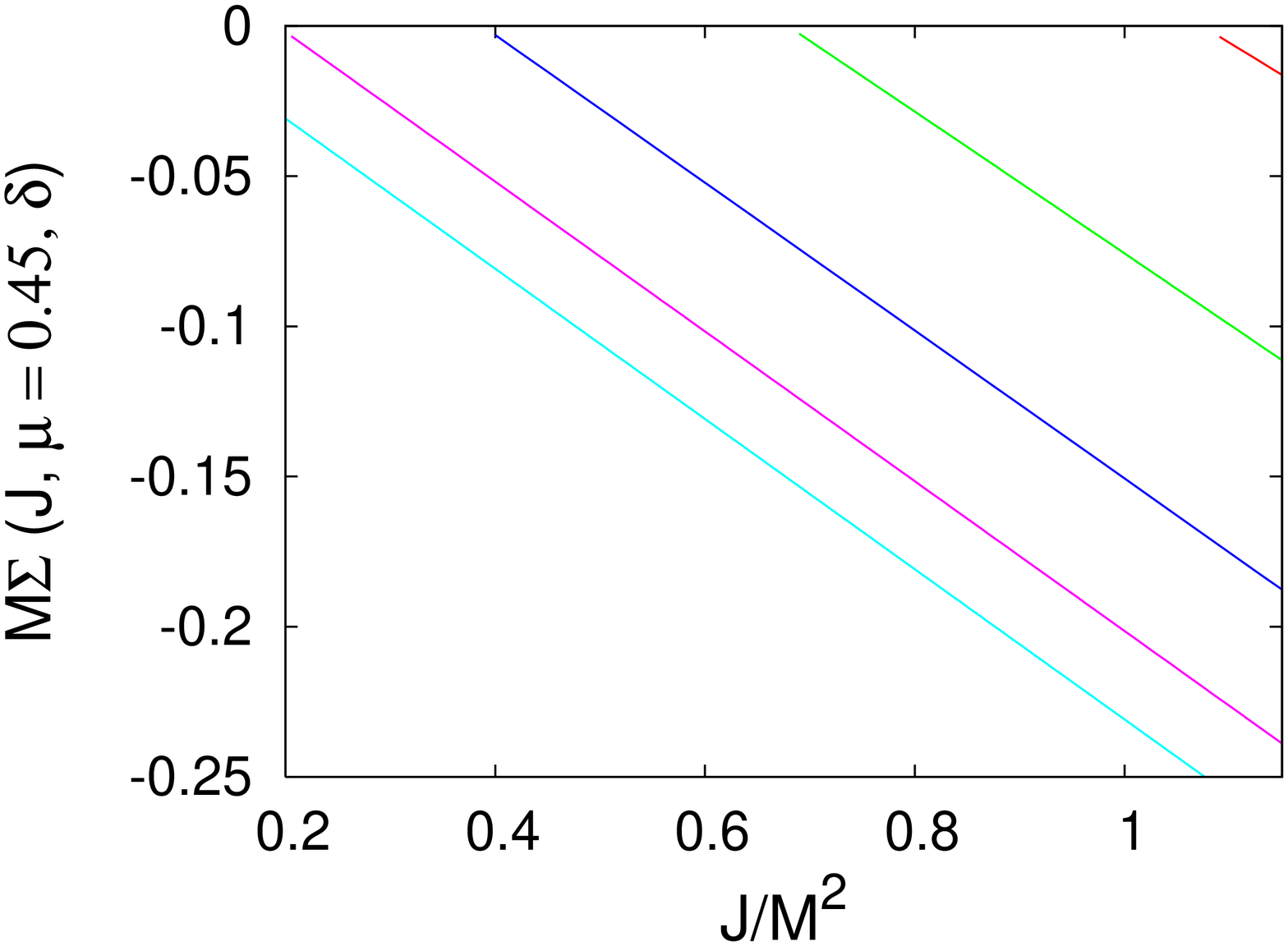}
  \caption{\textit{Top row, left panel:} Timescale $\tau$ of the
    instability (with $\ell = m = 1$) as a function of the angular
    momentum $J$ of the gravastar, for fixed thickness of the shell
    $\delta/M = 0.4$ and different compactnesses $\mu$ (from left to
    right $\mu = 0.47,\,0.45,\,0.44$ and $0.43$).  The horizontal line
    shows a timescale of $1{\rm s}$ for gravastars with $M =
    10M_{\astrosun}$. \textit{Top row, right panel:} Frequency
    $\Sigma$ of the instability as a function of $J$ for the same
    cases in the left panel. \textit{Bottom row:} The same as in the
    top row, but for gravastars with fixed compactness $\mu = 0.45$
    and varying thickness of the shell $\delta$ (from left to right
    $\delta/M = 0.6,\,0.5,\,0.4,\,0.3$ and $0.2$).}
  \label{fig_delta}
\end{figure*} 

A more complete picture of the real and imaginary parts of the
eigenfrequencies for our representative rotating gravastars is shown
in Fig.~\ref{fig_delta}, where we have extensively explored the
parameter space $(\mu,\delta,J)$. More specifically, the top row shows
in the left panel the timescale $\tau$ of the instability (with $\ell
= m = 1$) as a function of the angular momentum $J$ of the gravastar,
for fixed thickness of the shell $\delta/M = 0.4$ and different
compactnesses $\mu$ (from left to right $\mu = 0.47,\,0.45,\,0.44$ and
$0.43$).  As a reference the horizontal line shows a timescale of
$1{\rm s}$ for gravastars with $M = 10M_{\astrosun}$. Similarly, the
right panel of the top row gives the frequency $\Sigma$ as a function
of $J$ for the same cases in the left panel. The bottom row, on the
other hand, shows the same quantities as the top one, but for
gravastars with fixed compactness $\mu = 0.45$ and varying thickness
of the shell $\delta$ (from left to right $\delta/M =
0.6,\,0.5,\,0.4,\,0.3,\,$ and $0.2$). 

Note that the timescale for the growth of the instability increases
over-exponentially with decreasing angular momentum $J$, and this is
why a small spin down of the gravastar is sufficient to make it
``effectively stable'' even if it possesses an ergoregion and is
mathematically unstable. Note also that the frequency $\Sigma$
increases only linearly with decreasing $J$. Finally, to aid those
interested in reproducing our results and for code-testing purposes,
we have collected in Table~\ref{table2} some numerical values for the
first unstable mode relative to our reference gravastar (\cf
Table~\ref{table1}). Clearly, the values of $\tau$ are very sensitive
to $J$, but the big difference in $\tau$ seen between the gravastar
with $J/M^2 = 0.4$ and the one with $J/M^2 = 0.2$ is also due to the
fact that in the first case the lowest unstable mode is still
$\ell=m=1$, while in the second case it is already $\ell=m=5$.

The concluding remark of this Section should be one of caution. All of
our treatment is based on the slow-rotation approximation, yet we have
stretched it to compute also models with $\Omega/\Omega_K =
1$. Furthermore, while the WKB formulas we used were derived in the
high-$m$ limit, we have effectively used them also for very low values
of $m$ and found, in particular, that most rotating gravastars are
unstable already for $\ell=m=1$ mode, as a result of their high
compactness. Although our approach is not dissimilar to the one made
in related works~\cite{Comins,Cardoso}, it is important to underline
that we expect our estimates to be accurate only for slowly rotating
gravastars. On the other hand, we also believe that the qualitative (and
possibly quantitative) picture derived here will remain unchanged also
when a more sophisticated analysis is performed.

\begin{table}
\caption{\label{table2}Typical values of $J/M^2$, $r_a$, $r_b$, $r_c$,
  $\Sigma$ and $\tau$ for for the first unstable mode of the gravastar
  with $\mu = 0.45$ and $\delta/M = 0.4$. Note that all values but the
  last one refer to the $\ell=m=1$ mode.}
\begin{ruledtabular}
\begin{tabular}{lllllll}
$J/M^2$ & $r_a/M$ & $r_b/M$ & $r_c/M$ &
$M\Sigma$ & $\tau/M$ & unstable mode\\ 
\hline
1.2 & 1.24 & 2.048 & 4.31 & -0.20 & $1.24\times10^3$ & $\ell=m=1$ \\
1.0 & 1.24 & 2.052 & 5.76 & -0.15 & $2.35\times10^3$ & $\ell=m=1$ \\
0.8 & 1.24 & 2.056 & 8.88 & -0.10 & $5.96\times10^3$ & $\ell=m=1$ \\
0.6 & 1.25 & 2.061 & 18.1 & -0.052 & $2.73\times10^4$ & $\ell=m=1$ \\
0.4 & 1.25 & 2.066 & 350 & -0.0027 & $1.12\times10^7$ & $\ell=m=1$ \\
0.2 & 1.70 & 1.982 & 592 & -0.0017 & $1.19\times10^{30}$ & $\ell=m=5$ \\
\end{tabular}
\end{ruledtabular}
\end{table}

\section{Conclusions}
\label{sec:Conclusions}

Motivated by recent work on this subject~\cite{Cardoso}, we have
investigated the ergoregion instability in rotating gravastars,
exploring a large space of parameters and taking into account the
limits on the thickness of the matter shell and on the compactness.
While we confirm the results of Cardoso et al.~\cite{Cardoso} for the
models they have considered, we also draw two conclusions which are
less restrictive than theirs. Firstly, we find that models of rotating
gravastars without an ergoregion (and therefore stable) can be
constructed even for extreme rotation rates, namely for models with
$J/M^2 \sim 1$. Hence, not all rotating gravastars possess an
ergoregion. Secondly, because stable gravastar models with $J/M^2 \sim
1$ can be constructed, we conclude that not all ultra-compact
astrophysical objects rotating with $J/M^2 \sim 1$ must be black
holes.

Besides clarifying these two important aspects of rotating gravastars,
our analysis also helps to further constrain the properties of these
ultra-compact objects. Building on our initial work~\cite{Chirenti},
in fact, we have computed an additional constraint on the maximum
compactness of a gravastar which is to be stable to the ergoregion
instability. Such a maximum compactness is still much larger
than that of typical neutron stars but also smaller than that of
black holes with $J/M^2 \sim 1$. This should help in the
important effort of distinguishing (rotating) gravastars from
(rotating) black holes.

\begin{acknowledgments}

  It is a pleasure to thank Shin'ichirou Yoshida for numerous discussions
  and suggestions, and Vitor Cardoso for sharing some of his numerical
  results and for useful comments. CBMHC gratefully acknowledges the
  Alexander von Humboldt 
  Society for a postdoctoral fellowship; part of this work was supported
  in part by the DFG grant SFB/Transregio~7.

\end{acknowledgments}

\appendix
\section*{Appendix}
\setcounter{section}{1}
\label{sec:Appendix}

We review the derivation of the WKB formulas (\ref{BS}) and
(\ref{tau}) used in this paper, following closely the treatment given
in ref.~\cite{Comins}. We have chosen to include this Appendix in the
paper, even though the usual WKB approximation is standard in quantum
mechanics textbooks, because the case in question is somewhat more
complicated than the usual textbook exercises. The equation to be
solved is [\cf eq.~(\ref{scalar})]
\begin{equation}
\psi_{,rr} + m^2T(r,\Sigma)\psi = 0\,,
\label{ap:scalar}
\end{equation}
which has four different regions with distinct physical behavior, as
described in Section~\ref{sec:Perturb}. We are interested in finding
the purely outgoing modes of this equation. These are given as poles
of the scattering amplitude $S \equiv {C_{\rm out}}/{C_{\rm in}}$,
where $C_{\rm out}$ and $C_{\rm in}$ are functions of the complex
frequency $\sigma$ ($\Sigma=\sigma/m$ and we assume a
$e^{i\sigma t}$ dependence for $\psi$) and denote the amplitude
of the outgoing and incoming waves at infinity, respectively. In a
pole of $S$, we will have $C_{\rm in} = 0$ and $C_{\rm out} \ne
0$. For the modes with small imaginary part,
$\Re(\sigma)\gg\Im(\sigma)$ it will be sufficient to determine the
eigenvalue $\sigma$ approximately as a real frequency (on the real
axis). In this spirit we define an auxiliary function ${\bar S(\sigma)
  = [S(\sigma^*)]^*}$, such that if $S$ has a pole at $\sigma_{\rm
  p}$, then ${\bar S}$ has a pole at $\sigma_{\rm p}^*$, where (*)
denotes complex conjugation.

If we restrict ourselves now to $\sigma \in \mathbb{R}$, we will have
conservation of energy and 
\begin{equation}
|S| = 1 \, \Rightarrow \, S(\sigma)[S(\sigma)]^* = 1
\, \Rightarrow \, S(\sigma) = [{\bar
    S}(\sigma)]^{-1}\,,
\end{equation}
where the last result is obtained because $\sigma$ is real. As the
relation above is valid for all real values of $\sigma$, it is valid
everywhere $S$ and ${\bar S}$ are analytic functions. Therefore,
returning to the complex plane, we can state that if ${\bar S}$ has a
pole at $\sigma_{\rm p}^*$, then $S$ has a zero at $\sigma_{\rm p}^*$
(and still a pole at $\sigma_{\rm p}$, which we assume to be simple),
and can be written approximately as
\begin{equation}
S = e^{2i\delta_0}\frac{\sigma-\sigma_{\rm
    p}^*}{\sigma-\sigma_{\rm p}} \quad \Rightarrow \quad S =
    e^{2i\delta_0}\frac{\sigma-\sigma_{\rm r} -
    {\rm i}/\tau}{\sigma-\sigma_{\rm r} + {\rm i}/\tau}
\label{S_final}
\end{equation}

Now we turn to the specific form of the solutions of
eq.~(\ref{ap:scalar}) in the four different regions as discussed in
Sect.~\ref{sec:Perturb}: region I or inner ``forbidden'' region,
region II or ``allowed'' region, region III or ``potential barrier'',
region IV: $r>r_c$ or external ``allowed'' region (\cf
Fig.~\ref{fig_well}). The connection formulas for the wave function to
the left and right of a turning point can be found in many standard
textbooks on quantum mechanics (\eg~\cite{Merzbacher}) and amount to
\begin{eqnarray}
\label{connection1}
\frac{2}{\sqrt{k}} \cos \left( \int_x^a k dx -
  \frac{\pi}{4}\right) \rightarrow \frac{1}{\sqrt{\kappa}}\exp\left(
  -\int_a^x 
  \kappa dx\right)\\
\label{connection2}
\frac{2}{\sqrt{k}} \sin \left( \int_x^a k dx -
  \frac{\pi}{4}\right) \leftarrow -\frac{1}{\sqrt{\kappa}}\exp\left(
  \int_a^x 
  \kappa dx\right)
\end{eqnarray}
for a turning point $x=a$ to the right of the classical region, and
\begin{eqnarray}
\label{connection3}
  \frac{1}{\sqrt{\kappa}}\exp\left( -\int_x^b
  \kappa dx\right) \rightarrow \frac{2}{\sqrt{k}} \cos \left(
  \int_b^x k dx - \frac{\pi}{4}\right) \\ 
\label{connection4}
 -\frac{1}{\sqrt{\kappa}}\exp\left( \int_x^b
  \kappa dx\right)   \leftarrow \frac{2}{\sqrt{k}} \sin \left(
  \int_b^x k dx - \frac{\pi}{4}\right)
\end{eqnarray}
for a turning point $x=b$ to the left of the classical region. In this
notation, the equation to be solved is $d^2\psi/dx^2 + k^2(x)\psi =
0$ or $d^2\psi/dx^2 - \kappa^2(x)\psi = 0$, with $k^2\,,\kappa^2 >
0$. We will need this formulas in what follows, to connect the wave
function in the different regions. The form of the radial function
$\psi(r)$ in region I is determined by the regularity condition at the
center ($\psi$ must vanish at $r=0$),
\begin{equation}
\psi_{\rm I} = \frac{C_1}{r^{1/2}|T|^{1/4}} \exp\left( -m\int_r^{r_a}
  \sqrt{|T|}\,dr\right)\,. 
\end{equation}
Using now the connection formula (\ref{connection3}), we can obtain
the form of the radial wave function in region II (connecting through
the first turning point $r = r_a$)
\begin{eqnarray}
&&\hskip -1.0cm\psi_{\rm II} = \frac{C_1e^{{\rm i} \zeta}}{r^{1/2}T^{1/4}} 
\exp\left( {\rm i}\,m\int_{r_b}^{r} \sqrt{T}\,dr\right) + \nonumber \\
&& + \frac{C_1e^{-{\rm i} \zeta}}{r^{1/2}T^{1/4}} 
\exp\left( -{\rm i}\,m\int_{r_b}^{r} \sqrt{T}\,dr\right)\,,
\end{eqnarray}
where $\zeta \equiv  -m\int_{r_a}^{r_b}\sqrt{|T|}\,dr -
{\pi}/{4}$. We can now write the solution in regions III and IV as
\begin{eqnarray}
&&\hskip -1.0cm\psi_{\rm III} = \frac{C_2}{r^{1/2}|T|^{1/4}} 
\exp\left(-m\int_{r_b}^{r} \sqrt{|T|}\,dr\right) + \nonumber \\
&&+ \frac{C_3}{r^{1/2}|T|^{1/4}} 
\exp\left(m\int_{r_b}^{r} \sqrt{|T|}\,dr\right)\,,\\
&&\hskip -1.0cm\psi_{\rm IV} = \frac{C_4}{r^{1/2}T^{1/4}} 
\exp\left( {\rm i}\,m\int_{r_c}^{r} \sqrt{T}\,dr\right) + \nonumber \\
&&+ \frac{C_5}{r^{1/2}T^{1/4}} 
\exp\left( -{\rm i}\,m\int_{r_c}^{r} \sqrt{T}\,dr\right)\,.
\end{eqnarray}
Combining the expressions for $\psi_{\rm II}$, $\psi_{\rm III}$ and
$\psi_{\rm IV}$ with the connection formulas
(\ref{connection1})-(\ref{connection4}), we can relate the amplitudes
of the waves before and after crossing the potential barrier. This is
done as in a standard scattering exercise, demanding the continuity of
the wave function and its first derivative across the turning points
$r_b$ and $r_c$. The result, as stated in ref.~\cite{Merzbacher}, is
remarkably simple and best expressed in matrix notation:
\begin{equation}
\left(
\begin{array}{c}
C_1e^{{\rm i}\zeta} \\
C_1e^{-{\rm i}\zeta}
\end{array} \right) = \frac{1}{2}
\left(
\begin{array}{cc}
2\eta + \frac{1}{2\eta} & {\rm i}\left(2\eta - \frac{1}{2\eta}\right) \\
-{\rm i}\left(2\eta - \frac{1}{2\eta}\right) & 2\eta + \frac{1}{2\eta}
\end{array}
\right)
\left(
\begin{array}{c}
C_4 \\
C_5
\end{array} \right)\,,
\label{matrix}
\end{equation} 
where $\eta \equiv \exp\left( m \int_{r_b}^{r_c}\sqrt{|T|}\,dr \right)$.
We can now identify the amplitudes $C_4$ and $C_5$ with the amplitudes
$C_{\rm in}$ and $C_{\rm out}$ of the incoming and outgoing waves at
infinity that we have introduced in the definition of $S$. If $\Sigma$
is negative, the case in which we have unstable modes, then
$C_4=C_{\rm out}$ and $C_5=C_{\rm in}$. Inverting now
eq.~(\ref{matrix}) in order to obtain $C_4$ and $C_5$ in terms of
$C_1$, they can be substituted in the definition of $S$ to give
\begin{equation}
S= \frac{C_4}{C_5} = \frac{(4\eta^2+1)e^{{\rm i}\zeta} - 
{\rm i}(4\eta^2-1)e^{-{\rm i}\zeta}}{{\rm i}(4\eta^2-1)e^{{\rm i}\zeta} + 
(4\eta^2+1)e^{-{\rm i}\zeta}}\,.
\label{S_cont}
\end{equation}
As we have discussed above in the beginning of this Appendix, we are
interested in finding the purely outgoing modes, which are poles of
the scattering amplitude $S$. Also, we made the assumption that this
poles will occur for complex frequencies $\sigma_{\rm p}$ which lie on
the complex plane, but very close to the real axis. But the imaginary
part of $\sigma_{\rm p}$ will only be very small if the ``barrier
penetration'' integral $\eta$ is very large. Taking the limit of $S$
as $\eta \rightarrow \infty$, we have
\begin{equation}
S \simeq \frac{e^{{\rm i}\zeta} -
  {\rm i}e^{-{\rm i}\zeta}}{{\rm i}e^{{\rm i}\zeta} + e^{-{\rm i}\zeta}} =
  -{\rm i} \quad \textrm{for} \quad \eta \to \infty, 
\end{equation}
unless we have $e^{{\rm i}\zeta}-{\rm i}e^{-{\rm i}\zeta}=0$, in which
case we can write $S$ as 
\begin{eqnarray}
S  
 = {\rm i}\frac{-4\eta^2({\rm i}e^{{\rm i}\zeta}+e^{-{\rm i}\zeta}) - 
({\rm i}e^{{\rm i}\zeta}-e^{-{\rm i}\zeta})}
{4\eta^2({\rm i}e^{{\rm i}\zeta}+e^{-{\rm i}\zeta}) - 
({\rm i}e^{{\rm i}\zeta}-e^{-{\rm i}\zeta})} = {\rm i}\,.
\end{eqnarray}
Therefore we can see that $S$ will have a resonance at a frequency
$\sigma_n$ near the frequency for which
$e^{{\rm i}\zeta}-{\rm i}e^{-{\rm i}\zeta}=0$ and thus $\zeta = n\pi +
{\pi}/{4}$, with $n$ an integer.
We can now write $\zeta$ as a series expansion around $\sigma_n$,
$\zeta(\sigma) \simeq n\pi + {\pi}/{4} + \alpha_n(\sigma-\sigma_n)$,
where
\begin{equation}
\alpha_n = \frac{d}{d\sigma}\left(m
\int_{r_a}^{r_b}\sqrt{T}\,dr\right)\bigg|_{\sigma=\sigma_n}\,.
\end{equation}
Substituting now $\zeta(\sigma)$ into eq.~(\ref{S_cont}), we obtain at
first order in $(\sigma-\sigma_n)$ and dropping the index $n$ on
$\alpha_n$
\begin{eqnarray}
S &=& \frac{-\alpha(\sigma-\sigma_n) + 1/4\eta^2 + {\rm i}\left[
    \alpha(\sigma-\sigma_n) + 1/4\eta^2\right]}
{-\alpha(\sigma-\sigma_n) + 1/4\eta^2 - {\rm i}\left[
    \alpha(\sigma-\sigma_n) + 1/4\eta^2\right]}\,, \nonumber \\
\end{eqnarray}
which can then be rewritten in the form of eq.~(\ref{S_final}),
\begin{equation}
S = {\rm i} \frac{\sigma-\sigma_n-{\rm i}/4\eta^2\alpha}
{\sigma-\sigma_n+{\rm i}/4\eta^2\alpha}\,.
\label{S_tau}
\end{equation}
Combining the definition of $\zeta$ and its expansion and taking
 $\sigma=\sigma_n$ we finally obtain eq.~(\ref{BS}) and comparing
 eqs.~(\ref{S_final}) and (\ref{S_tau}), we obtain eq.~(\ref{tau}),
 both presented in Section~\ref{sec:Perturb}.

\end{document}